\documentclass[iop]{emulateapj}
\usepackage{ amssymb }
\usepackage{amsmath}
\usepackage{wasysym}
\usepackage{listings}
\usepackage{multirow}
\newcommand{\myemail}{drew.brisbin@mail.udp.cl}

\shorttitle{SWIMMERs: potential and limitations of propellantless interstellar travel}
\shortauthors{Brisbin, D.}

\begin{document}


\title{Spacecraft With Interstellar Medium Momentum Exchange Reactions:\\
 The potential and limitations of propellantless interstellar travel}


\author{Drew Brisbin}
\affil{\\
\textsc{Contact}\\
Universidad Diego Portales \\
N\'{u}cleo de Astronom\'{\i}a \\
Av. Ej\'{e}rcito 441, Santiago, Chile\\
\\
\textsc{Affiliation}\\
FONDECYT Postdoctorado \\
Universidad Diego Portales \\
N\'{u}cleo de Astronom\'{\i}a \\
\\
\myemail
}

\begin{abstract}
All known interstellar transportation methods encounter monumental technological or engineering roadblocks, or even rely on speculative unknown science. In particular light of the recent public excitement and ensuing disappointment regarding the exotic ``EM drive'' it is worthwhile to point out that propellantless space travel is eminently possible based on well established physical principles. Here a new mode of transport is proposed which relies on electric-field moderated momentum exchange with the ionized particles in the interstellar medium. The application of this mechanism faces significant challenges requiring industrial-scale exploitation of space but the technological roadblocks are different than those presented by light sails or particle beam powered craft, and may be more easily addressed depending on the uncertain march of technology. This mode of space travel is well suited to energy efficient travel at velocities $\lesssim 0.05c$ and compares exceptionally well to light sails on an energy expenditure basis.  It therefore represents an extremely attractive mode of transport for slow ($\sim$multi-century long) voyages carrying massive payloads to nearby stellar neighbors. This would be useful for missions that would otherwise be too energy intensive to carry out, such as transporting a generation ship or bulk materials for a future colony around  $\alpha$ Cen A.
\end{abstract}





\section{Introduction}\label{sec:intro}
The tyranny of the rocket equation has long been recognized as an impediment to becoming a truly spacefaring species. Due to the exorbitant reaction mass required for traditional rockets in interstellar travel, there has been considerable attention to methods of space travel that circumvent the rocket equation. Laser-driven light sails are a prominent and long-standing idea (see for example [1] and references therein). While light sails are well established and also the engines of the widely publicized ``Breakthrough Starshot'' program [2] and ``Project Dragonfly'' [3], their thrust is fundamentally limited to 6.67 N GW$^{-1}$. For comparison, the Three Gorges Dam, the largest capacity power plant currently in operation, has a capacity of about 22.5 GW. If this power was transmitted with perfect efficiency to a light sail it would provide thrust equivalent to the force required to lift a 15 kg mass on Earth.  Scaling light sails up to larger-than-gram-scale spacecraft therefore necessarily depends on humanity's ability to harness incredible power. Alternatively, direct sunlight could be used as a source of photon pressure. Unfortunately, the material properties suggested to be necessary for a practical interstellar solar sail require extremely low areal density materials with $\sigma \lesssim 10^{-3}$ g m$^{-2}$ [4]. Current state-of-the-art reflective films developed for light sails reach areal densities of $\sim$10 g m$^{-2}$, or four orders of magnitude too dense even without including any support structure or payload, so it is uncertain when if ever suitable materials will be developed for a solar sail [5].

Another idea using external reaction mass is the particle-beam powered spacecraft. This hinges on a sail formed by an extended electric or magnetic field which is able to deflect a remotely-beamed stream of charged particles. Since charged particles carry much more momentum per unit energy than photons this could have much lower power requirements than light sails. This concept has its origins in the Magsail, a large loop of current carrying wire which deflects passing charged particles in the interstellar medium (ISM), eliciting a drag force which could be used as a brake to slow spacecraft down to rest with respect to the ISM after a high speed journey [6]. To provide acceleration, one could simply replace the ISM with a beamed source of high velocity charged particles [7]. Providing a long distance beam of charged particles is, however, quite difficult because of beam divergence due to particle thermal motion, interaction with interplanetary or interstellar magnetic fields and electrostatic beam expansion in the case of non-neutral particle beams. Andrews (2003) suggests that it would be necessary to construct a highway of beam generators at least every AU or so along the route on which the craft accelerates [8]. The related concept of the electric sail instead uses an electric field generated by a positively charged grid of wires or wire spokes extending from a central hub to push against the outward streaming solar wind [9]. This concept has the near term potential to allow travel within our own stellar neighborhood with very low energy costs. The electric sail, like the Magsail however, ultimately relies on a drag force, decelerating the spacecraft to rest with respect to the surrounding medium (the outward moving solar wind in this case). It is therefore unable to accelerate beyond the heliosphere, nor can it accelerate directly inwards towards the sun while in the heliosphere (though tacking at an angle to the wind along with gravitational attraction do allow it to more slowly reduce its radial heliocentric distance).


It would be possible to overcome these obstacles by actively pushing against the charged particles of the ISM, rather than passively coming to rest with respect to the medium. These spacecraft with interstellar medium momentum exchange reactions (SWIMMERs) \textit{can} accelerate with respect to the ISM, are significantly more energy efficient than light sails, do not require pre-established infrastructure along the route and are based on elementary physical principles. Recently Robert Zubrin discussed his independent work on a ``dipole drive'' concept which bears a striking resemblance to the SWIMMER concept described here [10]. Although the two ideas are related and even share a similar geometry, they were arrived at independently. Furthermore the dipole drive, as described by Robert Zubrin, suffers from a flaw which prevents its successful acceleration in the ISM. The work presented here concerns the conceptual mechanism which allows SWIMMERs to accelerate through a stationary ISM. The mathematical properties of the idealized governing equations are developed and potential future SWIMMER missions are explored. This work adopts the nomenclature that $\textrm{Log(x)} \equiv \textrm{Log}_{10}\textrm{(x)}$.

\section{SWIMMER drive}\label{sec:1}
Both the Magsail and electric sail concepts rely on the fact that there is significant mass in the ISM (or the heliosphere) which can interact with relatively low mass structures consisting of charged or current carrying wires. How, then, could a spacecraft interact to accelerate rather than decelerate with respect to the surrounding medium?

Generally this will require a time varying electric field which can do work on the surrounding particles of the ISM. As a thought experiment, one can imagine a large paddy-wheel structure, a bit like the paddy-wheel of a paddle steamer boat, with two electric sails mounted opposite each other at the ends of two long tethers which are electrically connected and across which can be applied a potential difference. The tethers are mounted to a reaction wheel in the center and the whole system is set spinning with the axis perpendicular to the direction of travel (defined as the positive direction). If the spin rate is fast enough, there will be portions of the cycle during which the sails have negative velocity with respect to the ISM. As one of the electric sails (sail ``A'') approaches the portion of the cycle when its velocity is negative, a potential difference is applied, charging sail A positively and sail B negatively. In the frame of sail A, ions in the ISM are streaming towards it and pushing it in the desired direction of travel. Simultaneously the negatively charged sail B will be reflecting electrons in the positive direction causing some drag. Since ions out-mass electrons by at least a factor of $m_p$/$m_e$=1836, the electrons contribute negligible drag and can be ignored throughout the analysis. 
As the rotation cycle continues, sail A moves into the portion of the cycle where its velocity is no longer negative with respect to the ISM and the electric sails are neutralized. Then as sail B approaches the negative velocity portion of the cycle, the potential difference is turned on and reversed, charging sail B positively. By charging the sails positively only when they have negative velocities with respect to the ISM, they can operate like standard electric sails, exchanging momentum with the ambient medium and slowing down, while giving the overall spacecraft a positive momentum boost. In this way the positively charged electric sail pushes on the ambient ISM much like the submerged paddle on a paddle steamer boat pushes on the surrounding water. This process will also slow down the spin rate of the overall vehicle, but this can simply be spun up again by use of the central reaction wheel (using some on-board or beamed power source).  Although this is an illustrative example of a SWIMMER, it would likely pose extreme difficulties in implementation for fast (v$\sim$0.05 c) space travel. Note that each sail can only push on the ISM when its velocity is negative with respect to the ISM. To accelerate the spacecraft up to 0.05 c then, would require that the electric sails be moving at a speed of at at least 0.05 c with respect to the vehicle center of mass. Based on the requirement that the sails be as low mass as possible, likely constructed out of thin strands of superconducting wire, they are likely to be fragile, wispy structures and it seems unlikely that such sails and tethers could be robust against the strains involved while retaining their low mass.


\begin{figure}[!htb]
\centering
\includegraphics[width=0.45\textwidth,trim=0cm 0cm 0cm 0cm, clip=true,angle=0]{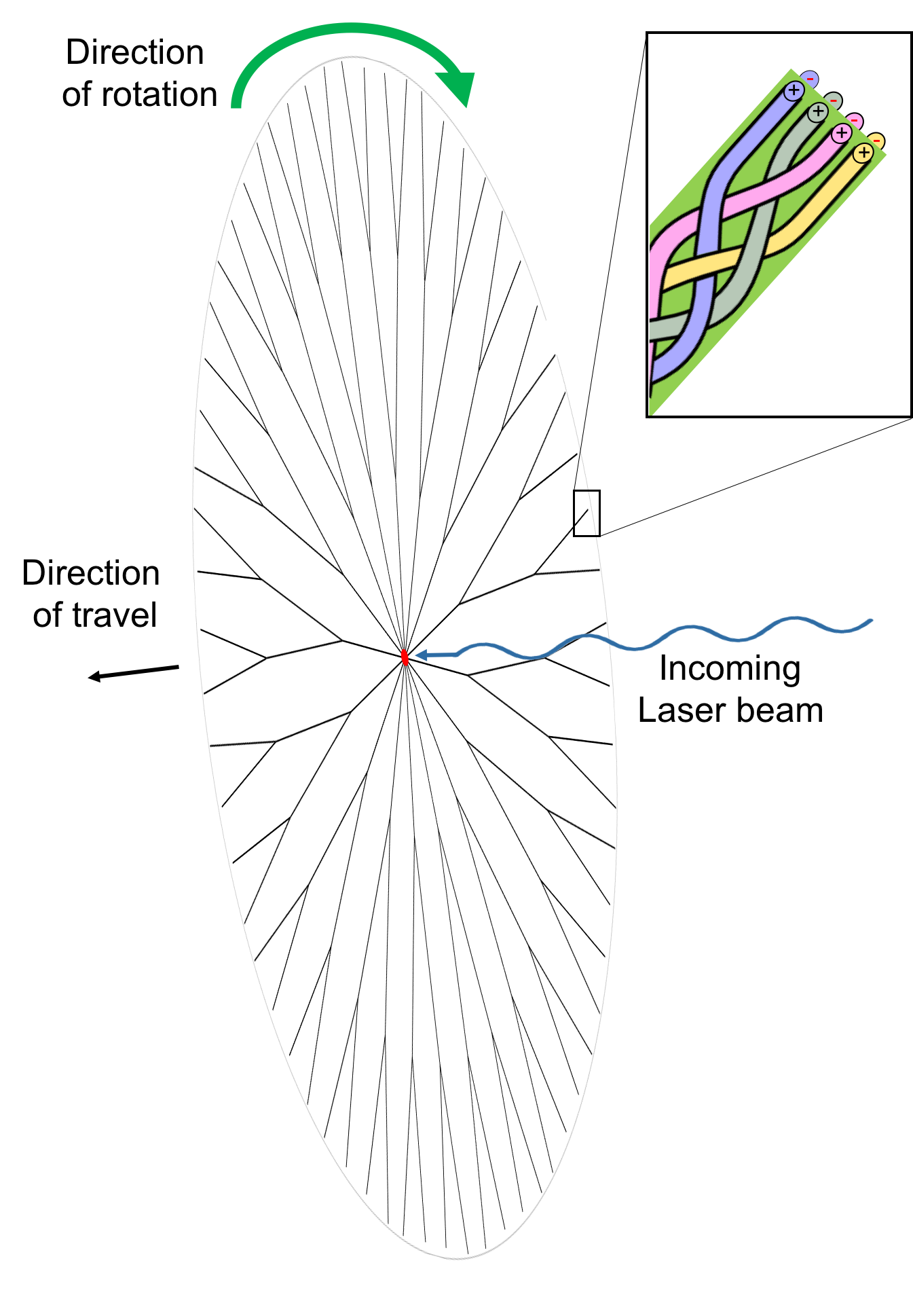}
	\caption{A schematic representation of a SWIMMER in operation. Tethers are shown branching off from each other. Laser light energy, represented by the blue squiggly arrow, is beamed from a power station located out of the figure to the right. It is absorbed and converted into electrical energy in the hub located at the center of the tether network, which would also be the location of the payload. The inset illustrates the bi-layer braided structure of a single tether. In its current charge state, with the front layer charged positively and the back layer negative, the SWIMMER is in a primer stage, pushing on positive ions in front to set up a clump in its immediate path. The braided inset figure is adapted from a creative commons file [11]. It has been cropped and edited for 3-d effect.
}
\label{fig:swimmerweb}
\end{figure}

Another option with fewer moving parts, could operate by first setting up inhomogeneities in the ISM arranged especially so that an electric field can push asymmetrically in one direction. There is an arbitrary number of configurations that could achieve this. One simple implementation, illustrated in Fig. \ref{fig:swimmerweb}, could feature a pusher plate made of a large grid or long tethers of wires moving face-on through the ISM (much like the proposed geometry of a standard electric sail). Unlike a standard electric sail, however, the grid of wires would actually be two identical layers of wire sandwiching a strong insulator between them to keep the two layers physically apart and electrically isolated. These wire grids or tethers could be made from very fine superconducting wire and the entire ensemble could be spun to create tension and keep the wire grids extended without heavy support structure. In the ``primer'' portion of the operation cycle, the front layer in the pusher plate is raised to a positive potential $\phi$, modestly above the stopping potential of the ions, $\phi_{\textrm{stop}}\equiv \frac{m_{\textrm{ion}}}{2 e} v_{\textrm{ion}}^2$ where $e$ represents the elementary charge, $m_{\textrm{ion}}$ is the mass of the ion species and $v_{\textrm{ion}}$ represents the maximum of either the ion thermal velocity or the streaming velocity (the velocity of the spacecraft itself, assuming travel through a  stationary medium). The back layer is charged to $-\phi$. Due to edge effects of the finite plates and the self-shielding behavior of plasmas, this results in a decaying electric potential of opposite sign on either side of the plates. Ions streaming towards the front positively charged layer slow down, building up an overdense clump in front of the pusher plate while an underdensity forms at the immediate location of the pusher plate. Then in the ``pull'' stage of the cycle the potential difference across the layers is reversed and significantly increased. The ion clump that was formed in front of the plate will be attracted to the negative front layer, pulling the spacecraft forward. Any ions that transition through the pusher plate at this moment will cause significant drag as they encounter the strongly rising electric potential crossing the thickness of the plate. Fortunately, the underdensity set up in the primer stage ensures there will be very few if any ions which will encounter the pusher plate before the clump. As the clump approaches the pusher plate, the potential difference is turned off and the clump is allowed to coast through the plate to the other side. In the final ``push'' stage the same potential difference is applied and the clump is further pushed backwards by the positive back layer of the pusher plate. The clump drifts away beyond the influence of the pusher plate and the cycle repeats. Fig. 2 shows the electric potential and ion density at various cycle stages for a simple model which represents the electric potential due to the pusher plate as two potential ramps extending out 20 m. The ions are assumed to initially be travelling rightward at 0.001 c, and their effect on the electric potential is ignored.

This process is conceptually straightforward and obeys all conservation laws. The spacecraft gains momentum by giving backward momentum to the ISM (pushing ions to the right in Fig. 2). The source of the potential difference does work in the primer stage when it sets up the positive potential, raising the electrical potential energy of the ion clump and again in the push stage when it raises the clump to a higher potential. In this idealized one-dimensional case, it is also very energy efficient. By appropriately tuning the cycle timing and the electrical potential levels the SWIMMER will avoid sending any ions initially in the vicinity of the plates forward to infinity (to the left). Electrons encountering the negative potentials will be reflected but this causes only a negligible momentum drag. In a real three dimensional case, there will also be loss of efficiency due to particles which do not interact perfectly in one dimension, but instead are pushed off to the side as they pass by the charged wires.

This qualitative conceptual analysis does not account for the self influencing behavior of plasmas. This will undoubtedly strongly influence the ion (and electron) distributions and the extended electric potential. Detailed particle-in-cell simulations will be necessary to investigate the optimal tuning of cycle timings and electrical potentials, and these will be affected by ISM density, pusher plate size, plasma (spacecraft) bulk velocity, plasma temperature and the available power. These simulations are beyond the scope of this paper but these collective effects are unlikely to eliminate the features created by the primer stage including a leading ion clump with an underdensity at the pusher plate. These inhomogeneities are the crucial feature that allow SWIMMERs to push on the ISM asymmetrically. Furthermore, this is only one possible configuration of a SWIMMER. It would also be possible to use multiple pusher plates to accelerate ion clumps across a series of potential differences to gain more thrust per ion, at the expense of a more complex and massive pusher plate. The effectiveness of a SWIMMER will ultimately need to be tested by simulation, small scale laboratory tests and real world application.

The configuration described here, a large pair of wire grids with opposite charges to push on the ambient ISM, is very similar to that recently described by Robert Zubrin as the dipole drive [10]. In the case of the dipole drive, however, the electric field is apparently static rather than pulsed, the wire grids are separated by a significant distance and they push on the charged particles as they pass between the plates. At first look, this seems like a reasonable and simpler approach. Two oppositely charged infinite plates produce a strong electric field between them and no electric field outside, so by simply pushing the heavy ions between the plates in the correct direction this static electric field should create thrust. Unfortunately, the approximation of infinite plates leads one astray here. In fact, a finite system of parallel plates will produce an electric field outside the plates pushing in the opposite direction. Although these fields will be weaker than the field between the plates, they will also extend over a larger region, cancelling out the thrust gained from particles between the plates. Indeed, \textit{any} system of charges over a finite area must leave the electric potential zero at infinity. Without any change in the electric field, any particles coming from far away and leaving far away begin and end with zero electric potential energy and no change to their kinetic energy. At most their velocity vector may change direction and lead to a change in momentum, but this change in momentum could only be used to decelerate (with respect to the ambient medium) or change direction.

\begin{figure}[!htb]
\centering
\includegraphics[width=0.5\textwidth,trim=0cm 0cm 0cm 0cm, clip=true,angle=0]{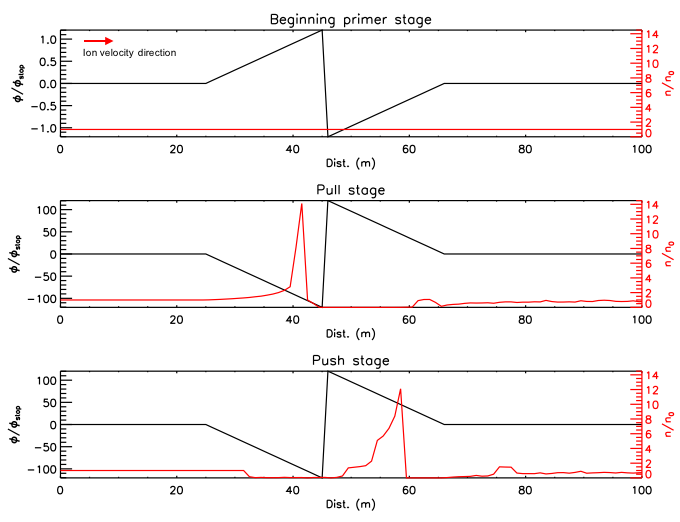}
	\caption{An example showing snapshots of the electrical potential and ion density in one dimension cutting across the SWIMMER pusher plate at various times.  The two-layered pusher plate is located at 45-46 m. Black lines indicate the electrical potential and red lines indicate the ion density (where the average ISM ion density is $n_0$). In the primer stage, a positive potential $\sim \phi_{\textrm{stop}}$ sets up an ion density gap at the location of the pusher plate and an overdensity on the upwind side. In the pull stage the potential difference is reversed and increased (note the change in the y-axis). This pull stage persists until just before the ion clump passes through the plate, at which point the pusher plate layers are neutralized. Once the clump passes through, the potential difference is restored beginning the push stage. This example assumes an initial uniform ion velocity of 0.001 c and does not account for the electric potential contributed by the ions or electrons.
	}
\label{fig:dipoleplate}
\end{figure}

\section{Mathematical expression of an idealized case}\label{sec:2}
To examine the limits of SWIMMERs, consider a spacecraft of mass, $m$, moving with some velocity, $v$, in the frame of the surrounding medium (this would be the stationary ISM frame in interstellar space or a frame that is comoving with the solar wind within a heliosphere). The spacecraft's direction of travel is defined to be positive. Ignoring the details of operation, at a given moment the spacecraft is able to inject some small amount of energy, $\Delta E$, into a collection of ions in its vicinity of mass $m_{\textrm{ion}}$, increasing the system's kinetic energy and changing the momentum of the spacecraft and the ions.

By conservation of energy and momentum we find the final ship velocity, $v'$, and the ion velocity $v_{\textrm{ion}}$' is:

\begin{multline}
v'=\frac{m^2 v}{m_{\textrm{ion}}m+m^2} \\
+\frac{(2\Delta E m_{\textrm{ion}}^2 m+2\Delta E m_{\textrm{ion}}m^2+m_{\textrm{ion}}^2 m^2 v^2)^{1/2}}{m_{\textrm{ion}}m+m^2}
\label{eq:vswimmer}
\end{multline}
\begin{multline*}
v_{\textrm{ion}}'=\frac{m v}{m_{\textrm{ion}}}-\frac{m^3 v}{m_{\textrm{ion}}(m_{\textrm{ion}}m+m^2)} \\
-\frac{m(m_{\textrm{ion}}m(2\Delta E (m_{\textrm{ion}}+m)+m_{\textrm{ion}}m v^2))^{1/2}}{m_{\textrm{ion}}(m_{\textrm{ion}}m+m^2)}
\end{multline*}

Now assume the energy donated to the ions is given by some power, $P$, applied over a small amount of time, $\Delta t$: $\Delta E=P\Delta t$. The mass of ions that the energy is applied to is given by the mass of ions swept out in time $\Delta t$ by some cross sectional interaction area of the pusher plate, $A$: $m_{\textrm{ion}}=A v\times n m_p$ where $n$ is the ion density and $m_p$ is used as the individual ion particle mass under the simplifying assumption that all the ions are protons. The acceleration of the spacecraft is found by making these substitutions in Eq. 1, subtracting the initial velocity $v$ from $v'$, dividing by $\Delta t$ and then taking the limit as $\Delta t \rightarrow0$. The force on the ship, $F_{\textrm{SWIMMER}}$, is simply acceleration times the ship mass and is given by:

\begin{equation}
F_{\textrm{SWIMMER}}=\pm(A m_p n v (2P+A m_p n v^3))^{1/2}-A m_p n v^2.
\label{eq:fswimmer}
\end{equation}

The argument of the square root is real and positive. Choosing the negative root corresponds to the situation in which the spacecraft gives up some momentum and sends the ions in the positive direction while slowing itself down, a braking force. Choosing the positive root corresponds to the spacecraft sending the ions in the negative direction and accelerating itself forward. A braking force could be generated by reversing the polarity of the pusher plate during the push and pull stages shown in Fig. 2. Eq. 2 represents the ideal limit of the force generated by any system generating thrust by pushing on the surrounding ions with perfect efficiency. 

The power referred to throughout this work is the delivered electrical power. Thus far the source of power for a SWIMMER has been ignored. There is no reason a SWIMMER could not use an onboard power source, making it totally independent of external infrastructure. This, of course, would require an exceptionally energy dense fuel source as well as a very efficient generator to achieve useful velocities for interstellar travel (note that if the spent fuel mass rate is high and used fuel is continuously ejected, this would alter eqs. 1 and 2 as the ship's mass decreases throughout the interaction, but if spent fuel is held on board these equations would not change). Beaming power remotely to the SWIMMER is possibly a more viable strategy for interstellar travel, which invites a direct comparison to light sails. In this case an additional $P/c$ term is included in eq. 2, corresponding to the photon pressure of the beamed energy being absorbed by the spacecraft. The total force is then:

\begin{equation}
F=\pm(A m_p n v (2P+A m_p n v^3))^{1/2}-A m_p n v^2\pm\frac{P}{c}.
\label{eq:fswimmer2}
\end{equation}

The photon force is added or subtracted depending on whether the beamed energy is directed in the same direction as the SWIMMER velocity or opposed to it respectively. Explicitly then, there are four different modes of operation for a SWIMMER depending on the orientation of the velocity, the photon force and the interaction force from pushing on the ions in the surrounding medium ($F_{\textrm{SWIMMER}}$). These are illustrated in Fig. 3 along with their corresponding implementations of eq. 3 with explicit positive and negative sign choices. These four modes of operation are referred to respectively as the  ``normal'' mode when the velocity, $F_{\textrm{SWIMMER}}$, and photon force are all oriented in the same direction; ``tractor beam'' mode when the velocity and $F_{\textrm{SWIMMER}}$ are aligned with each other but opposed to the photon force; ``destination braking'' when the photon force and velocity are aligned but opposed to $F_{\textrm{SWIMMER}}$; and ``home braking'' when the photon force and $F_{\textrm{SWIMMER}}$ are aligned but opposed to velocity. The ability to operate in these different modes is one of the advantages of SWIMMERs. Unlike light sails they are able to decelerate at their destinations or even be accelerated back towards their origin without additional infrastructure such as a laser array or giant reflector previously prepared at the destination or launched along with the spacecraft and allowed to travel ahead. Since $F_{\textrm{SWIMMER}}$ is velocity dependent and drops to zero at zero velocity, a SWIMMER would never be able to completely come to a stop and reverse in the dead of the ISM. Additional means of propulsion such as a modest rocket engine could be carried along to provide a small $\Delta$v to reverse the direction after most of the braking was accomplished in destination braking mode, and then the SWIMMER could begin operating again in tractor beam mode. If the destination were not the dead of the ISM, but a star with an active heliosphere, however, this would be unnecessary. Upon entering the alien heliosphere a SWIMMER could begin destination braking and reduce its velocity to near zero. Since the velocity referred to by eq. 2 is the spacecraft's velocity with respect to the interacting medium, within the alien heliosphere destination braking would allow the SWIMMER to approach the velocity of the outward streaming solar wind. The SWIMMER could then coast along with the solar wind until exiting the heliosphere, at which point its velocity with respect to the interacting medium would change from nearly zero to whatever the solar wind velocity was ($\sim$5$\times$10$^5$ m s$^{-1}$ around our sun and likely similar for other stars of similar type). At that point the SWIMMER could operate in tractor beam mode until approaching its origin. A SWIMMER launched within our heliosphere at velocities slower than the solar wind would operate in home braking mode, braking with respect to the solar wind but gaining velocity in the heliocentric frame. A SWIMMER which was sent to the edge of our heliosphere which needed to return swiftly could accelerate \textit{directly} inward in tractor beam mode, unlike electric sails.

\begin{figure}[!htb]
\centering
\includegraphics[width=0.45\textwidth,trim=0cm 0cm 0cm 0cm, clip=true,angle=0]{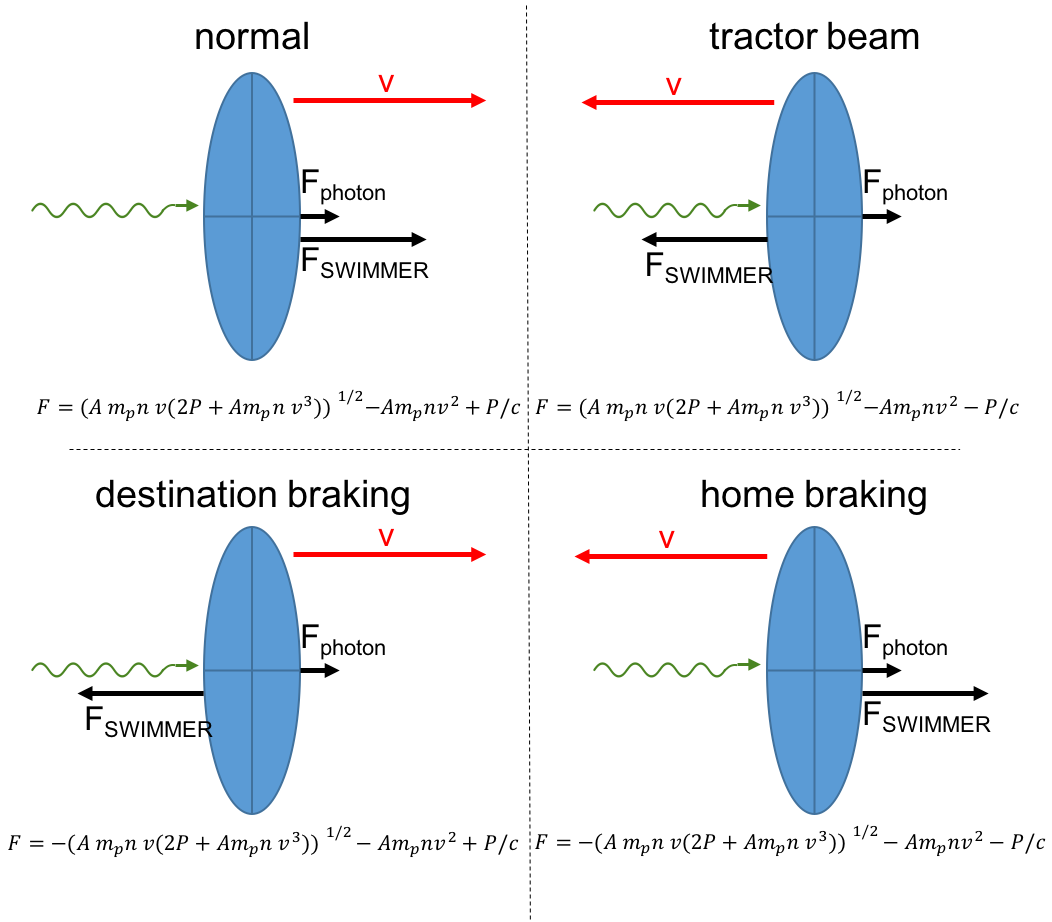}
	\caption{The four modes of operation for a SWIMMER. Red arrows indicates velocity, black arrows indicate the photon force and $F_{\textrm{SWIMMER}}$, green squiggle arrows indicates the direction of the energy beam. The equation for the total force in each mode is given at the bottom of each panel with explicit positive and negative signs. By definition velocity is in the positive direction (so in the two left panels positive is to the right, in the two right panels positive is to the left)}
\label{fig:operationmodes}
\end{figure}

All four of these modes of operation could be useful for different missions, however, for any SWIMMER mission to another stellar system, the normal mode will be used for the bulk of the journey. Therefore the mathematical description of normal mode, which uses the positive root and the positive photon force deserves further consideration. It will also be useful to consider the ratio, $R$, of the force on a SWIMMER in normal mode to the force on an ideal light sail with equal delivered power $(F=2 P/c)$, which can be written as:

\begin{equation}
R=\frac{1}{2}(1-\frac{A}{P}c ~m_p n v^2+c(2\frac{A}{P}m_p n v+(\frac{A}{P})^2 m_p^2 n^2 v^4)^{1/2}).
\label{eq:rswimmer}
\end{equation}

In Fig. 4 $R$ is shown as a function of velocity for a few values of $A/P$. There is some uncertainty surrounding the structure and properties of the local ISM, but there is general consensus that a journey to $\alpha$ Cen A will involve passage through some combination of the Local Interstellar Cloud, the Circum-Heliospheric Interstellar Medium and the G Cloud. Therefore, a conservatively low ion density of $n=0.07$cm$^{-3}$, consistent with the estimated densities in these clouds, is used in Fig. 4 (eg. [12]). Fig. 4 shows the force initially rising with velocity due to the increasing volume of ISM swept out. The force peaks at some velocity, $v_{\textrm{peak}}$, and then decreases as the ratio of the change in momentum to change in energy shrinks with faster ion velocities. Due to this initial rise in force with velocity, it may be useful to give SWIMMERs operating in normal mode an initial velocity boost through other means (such as conventional rockets, gravitational assists, particle beam assists or through home braking SWIMMER mode operations allowing electric sail-style passive interaction with the solar wind) to take advantage of the forces at higher velocities.

\begin{figure}[!htb]
\centering
\includegraphics[width=0.45\textwidth,trim=0cm 0cm 0cm 0cm, clip=true,angle=90]{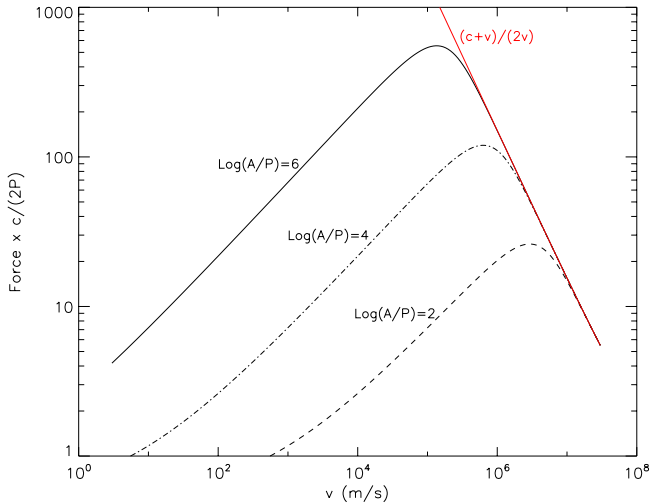}
	\caption{The total force on a SWIMMER in normal operation ($F_{\textrm{SWIMMER}} + F_{\textrm{photon}}$) divided by force on an ideal light sail (Eq. \ref{eq:rswimmer}), as a function of velocity. Trends are shown for Log($A/P$)=6, 4 and 2 in solid, dash dotted and dashed lines respectively. In general, high values of $A/P$ give superior performance relative to light sails. The straight red line indicates $(c+v)/(2v)$ the ratio approached at high velocities.		}
\label{fig:fofv}
\end{figure}

Taking the derivative of eq. 3 with respect to $v$ and setting it equal to zero gives the velocity of this peak force:

\begin{equation}
v_{\textrm{peak}}^3=\frac{P}{4 m_p n A}.
\label{eq:vpeak}
\end{equation}

Larger $A/P$ values give significantly better performance at lower velocities, but trend together as velocity increases, with the force approaching $\frac{P}{v}\frac{c+v}{c}$ (the ratio $R$ approaches $\frac{c+v}{2v}$, shown by the red line in Fig. 4). This high velocity limit implies an order of magnitude larger force for SWIMMERs relative to light sails up to $v=c/19$ or about 5\% c. This also indicates that it is not advantageous to increase the pusher plate area arbitrarily. If, for instance, a SWIMMER began operation with an initial velocity of 10$^6$ m s$^{-1}$ with Log($A/P$)=3, then increasing $A$ by a factor of 100 will not dramatically increase the total force, but significantly increasing the pusher plate area would presumably require an increase in the spacecraft mass and thus a net decrease to the acceleration. It might even be useful to adjust pusher plate area en route by discarding bits of the pusher plate as the SWIMMER reaches higher velocities. Also note that increasing $A/P$ by simply reducing the power will increase $R$ as seen in Fig. 4, but it decreases the overall force.

Finally, note that just as SWIMMERs should not have arbitrarily large pusher plates, they also should not be completely dominated by the mass of the power system which converts beamed light into electrical energy. If the SWIMMER's mass were dominated by the power conversion system, the mass grows with $P$. If the power system is considered as a thermodynamic heat engine, absorbing the remotely beamed energy and converting it to electrical power through a temperature differential with a heat sink radiating to empty space, then to convert more power at the same efficiency, the surface area of the heat sink, $S$, must increase proportional to $P$. Exactly how the mass of the heat sink increases with surface area depends on its geometry, but it is safe to assume $m\propto S^\gamma \propto P^{\gamma}$ where $1\leq \gamma \leq 3/2$. Considering the most optimistic case, $m\propto P$, the acceleration is given by dividing the force by $\kappa P$ (where $\kappa$ represents the specific power, W kg$^{-1}$ of the power conversion system). The first derivative of the acceleration with respect to $P$ in this case is always negative indicating that higher acceleration could be achieved by decreasing the power and the mass associated with the power system.

\section{Potential SWIMMER missions}\label{sec:3}
To illustrate the potential of SWIMMERs for interstellar travel, it is helpful to consider some possible future missions. The stationary infrastructure associated with the power beaming station has already been explored in some detail by other authors regarding light sails, and the associated strategies and technological obstacles for this component are equally applicable to SWIMMERs. Therefore the primary focus of this investigation is on the SWIMMER itself rather than its remote power source. In general, the specific power of the onboard SWIMMER power converter is an important parameter as it determines the mass devoted to the onboard power systems. Photovoltaic cells have fairly poor specific power, $\sim$80 W kg$^{-1}$ in space based applications [13]. Although technological advances such as inflatable solar arrays might improve their specific power, even these foreseen developments may not increase photovoltaic specific power sufficiently for use in a SWIMMER.  Rectennas may be more promising, with near-term estimates of specific power as high as 4 kW kg$^{-1}$ [13]. Although there is ongoing work to extend rectennas to the optical regime (eg. [14]), current rectennas are only able to convert light at $\sim$cm wavelengths to electrical power. At such long wavelengths however, the diffraction of the light beam would be too large to provide useful power at interstellar distances without an interstellar highway of booster beams or lenses along the route of travel. Perhaps the ideal power converter would be a simple reflector consisting of a thin aluminized membrane stretched across an aperture and electrostatically curved to a focus by a grid of charged wires behind it. Beamed optical or UV light could then hit the reflector and converge towards a focus at the hot side of a heat engine. The heat engine would need to be very low mass, but not necessarily efficient in terms of electrical output power to incident radiative power. In fact, an inefficient heat engine with a relatively hot ``cold'' side will radiate to space more efficiently and require a lower mass radiative heat sink, at the cost of requiring more energy output from the remote power beam station. The practical limits of such a system are not well known and estimating them is beyond the scope of this paper. Instead, it is assumed that a specific power of 4 kW kg$^{-1}$ is achievable. If heat engines are unable to achieve this, and neither rectennas nor photovoltaics are able to be sufficiently developed, then the achievable travel times will be longer than anticipated here. For the following examples a simplified model of the ISM and the heliospheres of the Sun and $\alpha$ Cen A is adopted. The ISM is assumed to be uniform with a density of $0.07$cm$^{-3}$, a temperature of 7000 K and therefore an electron Debye length $\lambda_D$=21.8 m. The heliospheres of both the Sun and $\alpha$ Cen A are assumed to have a density of $7.3$cm$^{-3}$, a temperature of 140000 K and therefore an electron Debye length $\lambda_D$=9.5 m. Furthermore, the solar wind in both heliospheres is assumed to be uniformly streaming outward at a velocity of 5$\times$10$^5$ m s$^{-1}$ out to a distance of 100 AU at which point the surrounding medium abruptly transitions to a stationary ISM. A summary of the mission parameters is shown in Table \ref{tab:missionparams}.

\begin{table}[h]
\caption{Mission parameters}
\label{tab:missionparams}
\centering
\begin{tabular}{r l l}
\hline\hline
  & Space probe & Ark ship \\ \hline
 V$_{\textrm{max}}$ & 0.020 c & 0.014 c \\
 t$_{\textrm{cruise}}$ & 260 years & 370 years \\
 t$_{\textrm{cruise}}$/t$_{\textrm{sail}}$ & 0.33 & 0.16 \\
 P$_{\textrm{delivered}}$ & 10 MW & 10000 GW \\
 M$_{\textrm{pay}}$ & 1000 kg & 8$\times$10$^9$ kg \\
 M$_{\textrm{power}}$ & 2500 kg & 1$\times$10$^9$ kg \\ \hline
 \multicolumn{3}{c}{Initial pusher plate parameters}\\ \hline
 Summed tether length & 4.1$\times$10$^9$ m & 2.0$\times$10$^{16}$ m \\
 M$_{\textrm{pusher}}$ & 7400 kg & 3.7$\times$10$^{10}$ kg \\ \hline
  \multicolumn{3}{c}{Final pusher plate parameters}\\ \hline
 Summed tether length & 2.9$\times$10$^8$ m & 1.0$\times$10$^{15}$ m \\
 M$_{\textrm{pusher}}$ & 520 kg & 1.9$\times$10$^{9}$ kg \\  
 \hline
\end{tabular} \\
\raggedright Mission parameters for the missions discussed in sections 4.1 and 4.2. The parameter t$_{\textrm{cruise}}$ represents the time to traverse 1 pc in the ISM, ignoring initial time spent in the sun's heliosphere as well as time spent decelerating near  $\alpha$ Cen A. The fraction t$_{\textrm{cruise}}$/t$_{\textrm{sail}}$ compares this cruise time to the time required for an ideal, massless light sail with an equal payload pushed by an equal amount of power. M$_{\textrm{pay}}$, M$_{\textrm{power}}$, and M$_{\textrm{pusher}}$ represent the mass devoted to the payload, the power conversion systems, and the pusher plate tethers respectively. Note that M$_{\textrm{pusher}}$ decreases throughout the journey as the SWIMMER discards pusher plate mass. Summed tether length gives the summed length of all the tethers used in the pusher plate. This can be converted to a pusher plate cross sectional area by way of eq. \ref{eq:area}.
\end{table}

\subsection{Space probe rendezvous at $\alpha$ Cen A}
A relatively lower mass SWIMMER mission might have the goal of transporting a modest space probe, $m_{\textrm{pay}}=1000$ kg to $\alpha$ Cen A and then decelerating to allow gravitational capture for a permanent orbital space telescope. A modest electrical power delivered to the SWIMMER of 10 MW is assumed.  The pusher plate will be made up of several long tethers. In practice these tethers will consist of very fine braided filaments to prevent failure due to micrometeoroid and interstellar dust collision, as described for the electric sail [9]. From a material mass standpoint these are considered to be single wires with an effective diameter of 30 $\mu$m. This is equivalent in material to eight filaments with diameters of about 10 $\mu$m. Strategically weak breakpoints in the tethers are included which can be activated by simply increasing the pusher plate spin rate such that the centripetal force exceeds the break point capacity. As the SWIMMER reaches higher velocities then, it may leave behind mass from the pusher plate. Given the pulsed nature of the SWIMMER electric field, the wire tethers should be made out of superconducting materials. A full analysis of the material requirements is beyond the scope of this paper, and it will depend on the necessary current density based on the geometry of the pusher plate as well as the timescale of the primer, pull and push stages. For this example MgB$_2$ is used to represent one possible wire material. MgB$_2$ is a well known super conductor with a density of $\rho$=2570 kg m$^{-3}$ and a high critical temperature (T$_c$=39 K) which should passively reach super conductivity beyond $\sim$5-50 AU depending on its surface emissivity. A single charged wire will interact with charged particles passing within $\sim\lambda_D$ on either side of it. The total cross sectional interaction area is given by 

\begin{equation}
A=L\times2\lambda_D
\label{eq:area}
\end{equation}

where $L$ is the summed length of all the tethers. This cross sectional interaction area is somewhat of an idealization as the Debye length does not represent a hard cut off where particles suddenly cease to be effected by an electric field, and in regions where tethers intersect, part of their cross sectional areas will overlap. Nonetheless it is a sufficient estimate for our rough calculations. The mass devoted to this pusher plate will be $m_{\textrm{pusher}}=\rho \times L\times \pi r_{\textrm{wire}}^2$

The total mass of the SWIMMER ship is comprised of $m_{\textrm{pay}}$=1000 kg, $m_{\textrm{power}}$=2500 kg (given by the 10 MW supplied electric power and its assumed specific power) and $m_{\textrm{pusher}}$. At the moment it is unclear how much mass to devote to $m_{\textrm{pusher}}$, however a mass of 7400 kg will be shown to be a useful choice. The mass for the tethers could be mined in situ from asteroids. This mass provides for a total summed tether length of 4.1$\times$10$^9$ m. While this is seemingly a very long tether, it does not in any way represent the spatial scale of the SWIMMER, as the pusher plate will be made up of of several thousand tethers, possibly splitting off from each other at greater radial distances. The summed length is merely a useful value for determining the total cross sectional area in plasmas of different temperatures and densities.


The SWIMMER begins at rest with respect to the sun near its creation site by the asteroid belt at 3 AU. Within the Sun's heliosphere the SWIMMER will be able to operate in home braking mode by producing a drag force with respect to the solar wind. The pusher plate tethers will not be super conducting in the inner solar system, but even while operating totally passively with $P=0$, a static charge on the plates will produce a significant drag force accelerating the SWIMMER towards the velocity of the solar wind. Based on the summed tether length, our SWIMMER will have a total cross sectional interaction area of 7.7$\times$10$^{10}$ m$^2$ within the heliosphere. Using simple code written in Interactive Data Language (IDL) (available upon request) the SWIMMER path is iteratively tracked according to eq. 3 while also introducing a gravitational attraction inwards toward the sun. After 1.5 years the SWIMMER enters the ISM at 100 AU with a velocity of 4.0$\times$10$^5$ m s$^{-1}$.

Upon entering interstellar space, the SWIMMER begins normal mode operations. Simultaneously the ion density drops and the cross sectional area of our tethers increases by a factor of $\lambda_{D\textrm{(ISM)}}/\lambda_{D\textrm{(helio)}}=2.3$. At this distance from the sun the SWIMMER tethers will be superconducting, and 10 MW of electrical power are supplied. The SWIMMER will also begin discarding mass from the pusher plate as it accelerates. The optimal rate to discard mass will change based on the specific details of any given mass distribution, power and journey length. Analysis of mass discard rate is not necessary for a conceptual understanding of the SWIMMER mission, but is investigated briefly for completeness. To consider this situation the problem can be parameterized by assuming that at any given moment, if the pusher plate mass is a significant fraction of the total mass, $m_{\textrm{pusher}}$/$m_{\textrm{tot}} > \chi$, mass will be discarded from the pusher plate until $A=\psi A_{\textrm{peak}}$ where $A_{\textrm{peak}}$ is the pusher plate area that corresponds to the $A/P$ value which lets the current SWIMMER velocity match $v_{\textrm{peak}}$. The parameters $\psi$ and $\chi$ were experimentally varied over a range of values to find the minimum travel time. In this example a minimum 1 pc travel time of 260 years is found with $\chi=0.13$, and $\psi=0.53$. The SWIMMER arrives with a velocity of 6.0$\times$10$^6$ m s$^{-1}$ (0.02 c). Without allowing the pusher plate mass to be discarded en route, the journey would take slightly longer at 340 years.  For comparison, an ideal light sail dominated by $m_{\textrm{pay}}=1000$ kg (i.e. ignoring the light sail mass and assuming perfect reflectivity) pushed with the same delivered power, would take 790 years to complete the same journey and it would not be able to stop at the destination without very complicated optics such as a detachable mirror that sails out ahead [1].

As the SWIMMER approaches $\alpha$ Cen A it begins destination braking.  This would begin in nearby interstellar space at a distance of $\sim$13000 AU from $\alpha$ Cen A. By this point the SWIMMER has significantly reduced the mass of its pusher plate to 520 kg, with a corresponding interaction area of 1.3$\times$10$^{10}$ m$^2$. After 23 years of braking in the ISM, the SWIMMER enters the $\alpha$ Cen A heliosphere at a distance of 100 AU and a velocity of 1.1$\times$10$^6$ m s$^{-1}$ relative to $\alpha$ Cen A. Again, the velocity of the surrounding medium changes, as does the density. The changed Debye length reduces the cross sectional interaction area of our tethers to 5.4$\times$10$^{9}$ m$^2$. Within the heliosphere gravitational attraction toward $\alpha$ Cen A is incorporated into the total force and power is reduced to zero. After a further 1.2 years of passive braking the SWIMMER reaches a distance of 2.8 AU with a velocity of 2.6$\times$10$^4$ m s$^{-1}$ with respect to $\alpha$ Cen A. At this distance the SWIMMER velocity is equivalent to the escape velocity and after another moment of braking the SWIMMER can either continue braking to eventually enter a circular orbit or stop braking and enter a highly elliptical orbit that will allow it to pass through the inner and outer regions of the $\alpha$ Cen A system. The full journey takes just under 290 years. This is a significant amount of time for a scientific endeavor, but there is good precedent for multi-century science projects for worthwhile investigations (c.f. [15-17]).


\subsection{Ark ship}
Due to their extremely favorable performance at lower power and velocities, SWIMMERs would make excellent transporters for large masses that can take long timescales. This could be used as the engine of a generation ship or perhaps a transporter for bulk colony materials sent out ahead of time before a fast moving, low mass, people transporter arrived. For this example assume a payload mass, $m_{\textrm{pay}}$=8$\times$10$^{9}$ kg, equivalent to the Super Orion ship discussed by Dyson (2002) [18]. Since such a mission would likely only be attempted after significant technological advances, slightly enhanced material properties are assumed including a specific power of the power conversion systems of 10 kW kg$^{-1}$, and superconducting materials which are able to passively operate beyond 3 AU. Delivered electrical power will be 10000 GW, thus $m_{\textrm{power}}$=1$\times$10$^{9}$ kg. With a pusher plate of mass of $m_{\textrm{pusher}}$=3.7$\times$10$^{10}$ kg the summed tether length is 2.0$\times$10$^{16}$ m. As before, this pusher plate mass is based on optimization of the travel time during the normal SWIMMER operation as a function of velocity, $\psi$, and $\chi$.

In the initial stage the ark SWIMMER accelerates in home braking mode from rest at 3 AU, with the full benefit of the beamed power. In the heliosphere the pusher plate has a cross sectional interaction area of 3.0$\times$10$^{17}$ m$^2$. Although it requires relatively little mass, this is, admittedly, very large ($\sim$20\% of the sun's cross sectional area). Care would need to be taken during construction to ensure tidal forces with any nearby asteroids do not disrupt the pusher plate.  This results in an eight-year journey to the edge of the heliosphere at 100 AU, where it enters the ISM at a velocity of 1.3$\times$10$^5$ m s$^{-1}$. Due to the larger Debye length, the cross sectional interaction area in the ISM is 6.8$\times$10$^{17}$ m$^2$.  Operating in normal mode with $\psi$=0.070 and $\chi$=0.17 it takes 370 years to travel 1 pc, at which point it has a velocity of 4.3$\times$10$^6$ m s$^{-1}$ and a remaining pusher plate mass of 1.9$\times$10$^{9}$ kg. For comparison, this 1 pc long journey through interstellar space would require 2300 years for an ideal light sail pushed by 10000 GW and with the same payload and negligible sail mass.

The very large pusher plate of the ark ship allows it to decelerate even faster than the previously considered space probe. If it begins destination braking at a distance of 6500 AU from $\alpha$ Cen A, then after 12 years it will reach the edge of the heliosphere with a velocity of 1.6$\times$10$^{6}$ m s$^{-1}$. Entering the heliosphere the cross sectional area changes as before and the SWIMMER continues destination braking with power. After another year the SWIMMER arrives at a a distance of 3.7 AU from the star and has braked to escape velocity at 2.3$\times$10$^{4}$ m s$^{-1}$ with respect to $\alpha$ Cen A. Slight variations in the onset of braking and the applied electrical power will allow it to reach any orbit within the heliosphere in comparable times. The full journey takes just under 400 years.

\section{Summary}\label{sec:4}
SWIMMERs represent a new mode of interstellar transport. By disposing of onboard reaction mass they circumvent the rocket equation, and by exchanging momentum with ions in the ISM they improve by orders of magnitude over the energy efficiency of traditional light sails at relatively low velocities. The key to this momentum exchange is the changing electric field which allows SWIMMERs to create inhomogeneities in the surrounding plasma and then push on these inhomogeneities to create thrust.  SWIMMERS perform exceptionally well at lower velocities, with their advantage over light sails diminishing quickly at v$>$0.05 c.  Furthermore, by relying on the ambient ISM as a momentum exchange medium, they are quite versatile, able to accelerate either away or towards a beamed energy source, opening up myriad opportunities to serve as one-way transport, roundtrips or even immobile statites hovering in stationary positions with respect to the Sun and serving as useful waypoints with infrastructure for other potential space transportation networks.

The examples discussed here only scratch the surface of the possible roles for SWIMMERs in our spacefaring future.  Their characteristics make them ideal for any mission with large masses in which relatively low velocities are acceptable. They are unlikely to be the sole mode of space transport due to their diminishing advantages at high velocities and their structural complexity which requires onboard power conversion systems with significant mass. Nonetheless, SWIMMERS will play an important role in future space exploration and augment other modes of transport. They might, for instance, also be well suited to aiding the construction of a fast interstellar highway by transporting massive particle beam stations along with their fuel supply out to stationary positions between us and our target destinations. These particle stations could be used to swiftly carry low mass Magsails along the path or augment the power of future SWIMMERs by replacing the stationary ISM with a corridor of fast moving beamed particles.

The missions analyzed here regard one way interstellar trips. While they do push the limits of current technology by assuming relatively high specific power electrical systems, very thin mass-produced super conducting wire, and low mass electrical insulators which can resist large potential differences (as well as very large laser array optics which are addressed in other works regarding light sails) there is no obvious material or theoretical limits which would prevent these missions from realization. Future work in this vein will need to examine several issues ignored here. Areas of further investigation, include the efficiency of the SWIMMER drive in three dimensions; the electrical potential and cycle timings during the pulsed SWIMMER operation and how they effect the required current density of the tethers; the expected impact of interstellar dust collisions and redundant tether configurations to avoid catastrophic damage from tether breakage and realistic limits on power conversion system capabilities.

As our understanding of interstellar travel develops, we must face the realization that, not only is it difficult, but there is no one-size-fits-all solution. Where SWIMMERs excel in one metric, other methods may excel in another. Ultimately our best strategy is to develop all possible methods in the hope that their synergy will provide a means to accomplish our goals.




\vspace{.1in}

\textsc{Acknowledgements} \\
Thanks to the the paper reviewer for insightful comments improving this work.
Drew Brisbin acknowledges support from FONDECYT postdoctorado project 3170974.

\vspace{.1in}
\textsc{References} 

\vspace{.1in}

1. R.L. Forward, ``Roundtrip Interstellar Travel Using Laser-Pushed Lightsails'', J. Spacecraft and Rockets, \textbf{21}, pp.187-195, 1984.

\vspace{.05in}

2. P. Lubin, ``A Roadmap to Interstellar Flight'', \textit{JBIS}, \textbf{69}, pp.40-72, 2016. 

\vspace{.05in}

3. N. Perakis, L.E. Schrenk, J. Gutsmiedl, A. Kroop, M.J. Losekamm, ``Project Dragonfly: A feasibility study of interstellar travel using laser-powered light sail propulsion'', Acta Astronautica, \textbf{129}, pp.316-324, 2016.

\vspace{.05in}

4. R. Heller, and M. Hippke, ``Deceleration of High-velocity Interstellar Photon Sails into Bound Orbits at  $\alpha$ Centauri'', Astrophysical Journal Letters, \textbf{835}, pp.L32, 2017.

\vspace{.05in}

5. D. Spieth, and R.M. Zubrin, ``Ultra-Thin Solar Sails for Interstellar Travel--Phase I Final Report'', NASA Institute for Advanced Concepts, Pioneer Astronautics Inc, 1999.

\vspace{.05in}

6. D.G. Andrews, and R.M. Zubrin, ``Magnetic sails and interstellar travel'', \textit{JBIS}, \textbf{43}, pp.265-272, 1990.

\vspace{.05in}

7. G.A. Landis, ``Interstellar flight by particle beam'', in \textit{AIP Conference Proceedings}, \textbf{vol. 552}, pp.393-396, 2001.

\vspace{.05in}

8. D.G. Andrews, ``Interstellar Transportation using Today's Physics'', Conference proceedings, American Institute of Aeronautics and Astronautics, \textbf{4691}, 2003.

\vspace{.05in}

9. P. Janhunen, ``Electric sail for spacecraft propulsion'', J. of Propulsion and Power, \textbf{20}, pp.763-764, 2004.

\vspace{.05in}

10. R.M. Zubrin, ``Dipole Drive for Space Propulsion'', Presented at \textit{Breakthrough Initiatives} conference, Stanford, California, April 2018.

\vspace{.05in}

11. Stilfehler, ``technique of 4 strand braiding'', Wikimedia Commons file (licensed for sharing and adaptation), https://commons.wikimedia.org/wiki/File:4\_Strand\_Braiding.png, last accessed on 18 March 2019.

\vspace{.05in}

12. I.A. Crawford, ``Project Icarus: A review of local interstellar medium properties of relevance for space missions to the nearest stars'', Acta Astronautica, \textbf{68}, pp.691-699, 2011.

\vspace{.05in}

13. J. Dankanich, C. Vassallo, and M. Tadge, ``Space-to-space power beaming enabling high performance rapid geocentric orbit transfer'', in \textit{51st AIAA/SAE/ASEE Joint Propulsion Conference}, 2015.

\vspace{.05in}

14. G. Moddel, and S. Grover (eds), \textit{``Rectenna solar cells''}, Springer, New York, NY, 2013.

\vspace{.05in}

15. A. Kivilaan, and R.S. Bandurski, ``The one hundred-year period for Dr. Beal's seed viability experiment'', American Journal of Botany, \textbf{68}, pp.1290-1292, 1981.

\vspace{.05in}

16. R. Johnston, ``World's slowest-moving drop caught on camera at last'', Nature News, \textbf{18}, 2013.

\vspace{.05in}

17. C. Cockell, ``The 500-year microbiology experiment'', Microbiology Today, \textbf{95}, pp.95-96, May 2014.

\vspace{.05in}

18. G. Dyson, \textit{``Project Orion: The True Story of the Atomic Spaceship''}, Henry Holt and co., New York, NY, 2002.

\end{document}